\documentclass{article}

\usepackage{arxiv}

\usepackage[utf8]{inputenc} 
\usepackage[T1]{fontenc}    
\usepackage{hyperref}       
\usepackage{url}            
\usepackage{booktabs}       
\usepackage{amsfonts}       
\usepackage{nicefrac}       
\usepackage{microtype}      

\usepackage{amsmath,amssymb}
\usepackage{graphicx}
\usepackage{textcomp}
\usepackage{xcolor}

\usepackage{mathtools}

\usepackage{algorithm,algpseudocode}
\makeatletter
\newcommand{\algmargin}{\the\ALG@thistlm}
\makeatother
\algnewcommand{\parState}[1]{\State%
    \parbox[t]{\dimexpr\linewidth-\algmargin}{\strut\hangindent=\algorithmicindent \hangafter=1 #1\strut}}
    
\DeclarePairedDelimiter\abs{\lvert}{\rvert}%
\DeclarePairedDelimiter\norm{\lVert}{\rVert}%

\makeatletter
\let\oldabs\abs
\def\abs{\@ifstar{\oldabs}{\oldabs*}}
\let\oldnorm\norm
\def\norm{\@ifstar{\oldnorm}{\oldnorm*}}
\makeatother

\title{Comparative Study between Adversarial Networks and Classical Techniques for Speech Enhancement}

\author{
  Tito Spadini\thanks{https://spadini.info} \\
  CECS\\
  Universidade Federal do ABC\\
  Santo André, SP \\
  \texttt{tito.caco@ufabc.edu.br} \\
   \And
 Ricardo Suyama \\
  CECS\\
  Universidade Federal do ABC\\
  Santo André, SP \\
  \texttt{ricardo.suyama@ufabc.edu.br} \\
}

\begin{document}
\maketitle

\begin{abstract}
Speech enhancement is a crucial task for several applications. Among the most explored techniques are the Wiener filter and the LogMMSE, but approaches exploring deep learning adapted to this task, such as SEGAN, have presented relevant results. This study compared the performance of the mentioned techniques in 85 noise conditions regarding quality, intelligibility, and distortion; and concluded that classical techniques continue to exhibit superior results for most scenarios, but, in severe noise scenarios, SEGAN performed better and with lower variance.
\end{abstract}


\section{Introduction}

Since the 1980s, speech enhancement and denoising researches exploit neural networks' ability to work as non-linear filters \cite{DNNSPEECHI, DNNSPEECHII, DNNSPEECHIII}, but it's performance was often unsatisfactory --- usually due to the reduced amount of training data or even the limited flexibility of the networks, caused by the inefficiency of the training algorithm for more extensive networks, with more neurons and layers. A new perspective for the use of neural networks arose after \cite{HINTON2006}, which has exhibited that employing an unsupervised pre-training of the network for layers can bypass the limitation found by gradient-based algorithms. This new possibility, coupled with the rise of new tools for training neural nets using GPU (Graphics Processing Unit) \cite{scherer2010evaluation}, has rekindled interest in neural networks.

Implementations of neural networks with large numbers of neurons and layers now include the term ``\emph {Deep}'' in their nomenclature, and therefore the term Deep Learning has become popular when referring to Deep Neural Networks (DNN). The flexibility and power of this type of neural network have shown promising results and have attracted a growing number of researchers in the field. However, there is a wide variety of DNN architectures; each with its respective characteristics and particularities, which may be more convenient for specific applications and therefore less suitable for others. For the applications discussed in this paper, an architecture known as convolutional autoencoder will be explored.

A recent approach called Generative Adversarial Networks (GAN) \cite{NIPS2014_5423} is a structure composed of individuals --- usually, two neural networks --- competing against each other and exploring concepts of Game Theory and Deep Learning. In this competitive two-player game, there is a well-prepared dataset, composed of samples of the same type, appropriately chosen, but with different attribute values. The \textit{Discriminator} player, $D$, has the purpose of discriminating whether a sample came from the original dataset or not; the \textit{Generator}, $G$, must capture the distribution of the original dataset and use it to generate entirely new samples. Thus, while one of the players intends to generate the perfect imitation of the original data, the other player tries to be the best possible counterfeit identifier.

For GAN-based methodes, the network training should occur gradually and concomitantly for both players, otherwise, an evolutionary imbalance may occur in favor of one of the actors, therefore, instead of achieving a good evolution for both, only one network will evolve minimally compared to the other, which does not even guarantee good results for at least one of the players. The GAN adjusting criterion is given by:

\begin{equation}
    \min_{G} \max_{D} V\left(D, G\right) = \mathbb{E}_{\pmb{x}\thicksim p_{\text{data}}\left(\pmb{x}\right)}\left[\log D\left(\pmb{x}\right)\right] + \mathbb{E}_{z\thicksim p_{z}\left(z\right)}\left[\log \left(1 - D\left(G\left(z\right)\right)\right)\right],
    \label{eq:minimax}
\end{equation}
where $V$ is the value function; $D$ is the Discriminator (player), a multi-layer perceptron that generates the probability that $\pmb{x}$ has originated from the legitimate data instead of the distribution $p_{G}$; $G$ is the Generator (player); $p_{\text{data}}$ is the data distribution; and $p_{\text{\pmb{z}}}$ is a prior in the input noise variables.

In the same way that GAN can learn a generative model for training data, conditional GAN (cGAN) \cite{CGAN}, as its name may suggest, provides a conditional generative model for the data. The data generation is based on a prior distribution and also on an additional input $\pmb{x}_c$, thus conditioning the distribution of the generated data to the additional information provided by $\pmb{x}_c$. The cost function of cGAN is given by:

\begin{equation}
    \min_{G} \max_{D} V\left(D, G\right) = \mathbb{E}_{\pmb{x}, \pmb{x}_c\thicksim p_{\text{data}}\left(\pmb{x}, \pmb{x}_c\right)}\left[\log D\left(\pmb{x},\pmb{x}_c\right)\right] + \mathbb{E}_{z\thicksim p_{z}\left(z\right), \pmb{x}_c\thicksim p_{data}\left(\pmb{x}_c\right)}\left[\log \left(1 - D\left(G\left(z, \pmb{x}_c\right),\pmb{x}_c\right)\right)\right],
    \label{eq:minimax_cGAN}
\end{equation}

Despite the advancement of the original GAN's and cGANs based on minimizing \eqref{eq:minimax} and \eqref{eq:minimax_cGAN}, in some cases the training may converge to solutions with performance below desired. For this reason, in \cite{LSGAN}, an alternative proposal was presented, called Least Squares GAN (LSGAN), which seeks to adapt the discriminator and the generator according to the following criteria:

\begin{equation}
    \min_{D} V_{LSGAN} \left(D\right) = \frac{1}{2}\mathbb{E}_{\pmb{x}, \pmb{x}_c\thicksim p_{\text{data}}\left(\pmb{x}, \pmb{x}_c\right)}\left[\left( D\left(\pmb{x},\pmb{x}_c\right)-1\right)^2\right] + \frac{1}{2}\mathbb{E}_{z\thicksim p_{z}\left(z\right), \pmb{x}_c\thicksim p_{data}\left(\pmb{x}_c\right)}\left[ D\left(G\left(z, \pmb{x}_c\right),\pmb{x}_c\right)^2\right],
    \label{eq:LSGAN_D}
\end{equation}

\begin{equation}
    \min_{G} V_{LSGAN} \left(D\right) = \frac{1}{2}\mathbb{E}_{z\thicksim p_{z}\left(z\right), \pmb{x}_c\thicksim p_{data}\left(\pmb{x}_c\right)}\left[ \left(D\left(G\left(z, \pmb{x}_c\right),\pmb{x}_c\right)-1\right)^2\right],
    \label{eq:LSGAN_G}
\end{equation}

A few years later, a GAN-based approach called Speech Enhancement GAN (SEGAN) was introduced in \cite{SEGAN}, exploring autoencoders neural networks adapted using RMSProp. Also, it has presented promising results. Further details on SEGAN will be provided in Section II.

One of the most decisive aspects regarding the adoption of an autoencoder, which in this case is fully convolutional, is its infundibuliform architecture capable of preserving the signal structure, ensuring that the obtained output will respect the same form used as network input. Also, the autoencoder can discard dispensable parts of the signals, i.e., noise, which causes the preservation of signal information to occur. However, due to the connections between encoding layers and decoding layers, the denoising effect caused by the network does not occur aggressively to the point of destroying the signal in terms of quality and intelligibility only for if noise reduction is achieved.

This work's objective is to compare SEGAN's performance against Wiener filter and Log Minimum Mean Square Error (LogMMSE), concerning quality and intelligibility through objective and perceptual metrics, for several different noise scenarios. However, different from \cite{SEGAN}, this work considered LogMMSE in its comparisons, in addition to Wiener's filter and SEGAN itself. Moreover, in addition to the Perceptual Evaluation of Speech Quality (PESQ) metric, explored in both studies, this work includes the Short-Time Objective Intelligibility (STOI) and Signal-to-Distortion Ratio (SDR) metrics. This work considered 5 different SNR scenarios (0 dB, 5 dB, 10 dB, 15 dB and 20 dB), which includes a new SNR value (20 dB) compared to those used in \cite{SEGAN}. The scenarios used in this paper considered 17 different types of noise, while \cite{SEGAN} used 5 types of noise.

In order to present the results of this study, the article adopts the following structure: section II presents a review on the main concepts related to SEGAN; section III covers the simulation scenarios and the metrics; section IV discuss the results and offers the final comments on the study; and section V exhibits the conclusions of the work.

\section{Speech Enhancement GAN - SEGAN}

\begin{figure}[htp]
\begin{center}
\includegraphics[width=.375\linewidth]{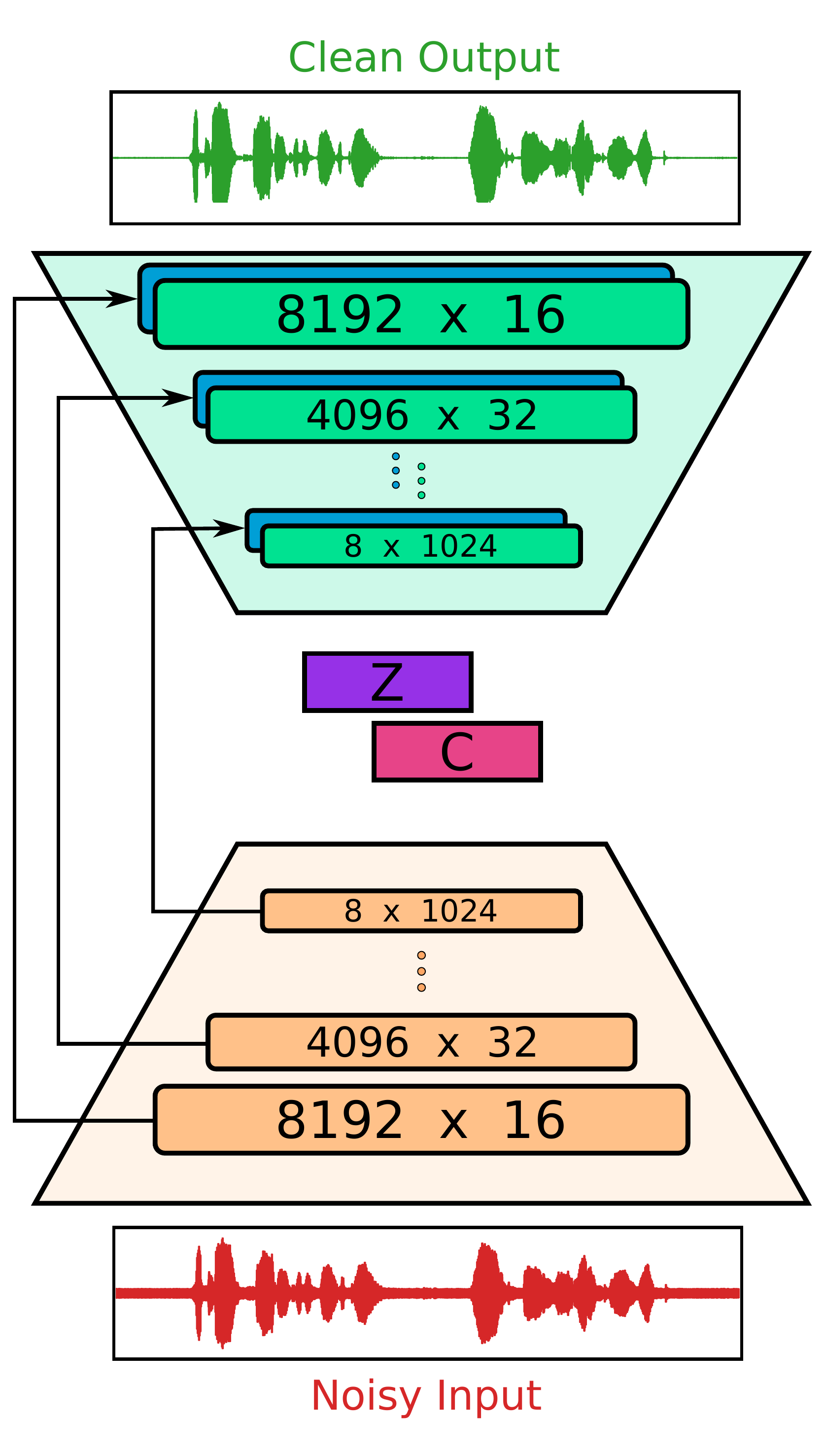}
\end{center}
\caption{\label{fig:autoencoder}Autoencoder architecture for speech enhancement. Based on \cite{SEGAN}.}
\end{figure}

\emph{Speech Enhancement Generative Adversarial Network} (SEGAN) \cite{SEGAN} is based on the idea introduced by cGAN, discussed in the previous section. The proposed structure for the generator, in this case, resembles an \emph{autoencoder}, shown in Figure \ref{fig:autoencoder}. The successive convolutional layers provide, at the end of the coding process, a vector $\mathbf{c}$ which corresponds to a condensed representation of the input signal. In this process, \emph{strided convolutions} layers are used, since this type of layer has been shown to be more stable in GAN training.

However, unlike the structure of a traditional autoencoder, the decoding process is done from the vector $\mathbf{c}$ concatenated with the vector of latent variables $\mathbf{z} $. This new vector is then subjected to a sequence of layers that seek to reverse the coding process by means of \emph{fractional-strided transposed convolutions} \cite{SEGAN}. The network structure also includes \emph{skip connections} connecting the outputs of the layers in the encoding process directly to corresponding layers in the decoding process. The reason for this was to try to maintain the underlying structure of the observed (noisy) data in the data being generated by GAN.

Although the presented LSGAN criterion may help with some known problems of structure adaptation, the authors propose, based on preliminary simulations, that the cost function includes an additional term in order to favor solutions that minimize the distance between generated data and authentic examples. The distance, however, is measured with the norm $\ell_1$, so that the adopted criterion is defined by:

\begin{equation}
    \min_{G} V_{LSGAN} \left(G\right) = \frac{1}{2}\mathbb{E}_{\pmb{z}\thicksim p_{\pmb{z}}\left(\pmb{z}\right),\ \pmb{\tilde{x}}\thicksim p_{data}\left(\pmb{\tilde{x}}\right)} \left[\left(D\left(G\left(\pmb{z},\  \pmb{\tilde{x}}\right),\ \pmb{\tilde{x}}\right)-1\right)^{2}\right] + \lambda\  \norm{G\left(\pmb{z},\ \pmb{\tilde{x}}\right)-\pmb{x}}_{1},
    \label{eq:LSGAN}
\end{equation}
where $\pmb{\tilde{x}}$ represents the input (noisy) signals.

\section{Simulation Setups}

To evaluate the performance, several testing scenarios were created from the combination of 20 voices of different people from the dataset VCTK-Corpus \cite{veaux2017cstr} reading three different sentences; (0 dB, 5 dB, 10 dB, 15 dB and 20 dB) for each of the 17 types of noise, all coming from the DEMAND \cite{thiemann2013demand} dataset, resulting in 85 different noise conditions for each sentence read by each person. In order to preserve a variability that would assist in the quest for broader generalizability in the SEGAN model training, the selected corpus was chosen to maintain a uniform gender distribution and to ensure different accents. Although the datasets were used in SEGAN's work, the selected individuals were not the same and more types of noise were used, including an extra SNR level (20 dB) that was not explored in the mentioned work.

With all different voice and noise scenarios previously detailed, 5100 different mixtures were obtained to be processed by the three speech enhancement techniques (Wiener filter, LogMMSE, and SEGAN). Using SoX, pre-processing was performed to ensure that all input signals conform to the 16 kHz, 16-bit, and mono configuration in WAV format. The selected Wiener filter belongs to the Scipy library; LogMMSE \cite{LOGMMSE, LOGMMSE_Python}, is also available as a Python package of the same name; and the SEGAN (pre-trained) model is the same as the original work \cite{SEGAN}, which is openly distributed by the SEGAN authors themselves in their GitHub repository. Such a model had been trained for 86 times using RMSprop \cite{tieleman2012lecture} and learning rate of 0.0002 in batches of size 400 \cite{SEGAN}. 

For each technique, after processing, an enhanced signal was obtained for each noisy signal used as input; and, based on the improved signal and the reference clean voice signal (directly from the VCTK-Corpus dataset), the (PESQ) \cite{PESQ}, a perceptual quality metric with values from -0.5 to 4.5; the STOI \cite{STOI}, which measures the improvement of intelligibility with values from 0 to 1; and the SDR \cite{SDR}, which quantifies the rate between the speech signal and the distorting effects of improved speech signal, were calculated to perform the improvement evaluation.

\section{Results}

\begin{figure}[htp]
\begin{minipage}[c]{0.475\linewidth}
\includegraphics[width=\linewidth]{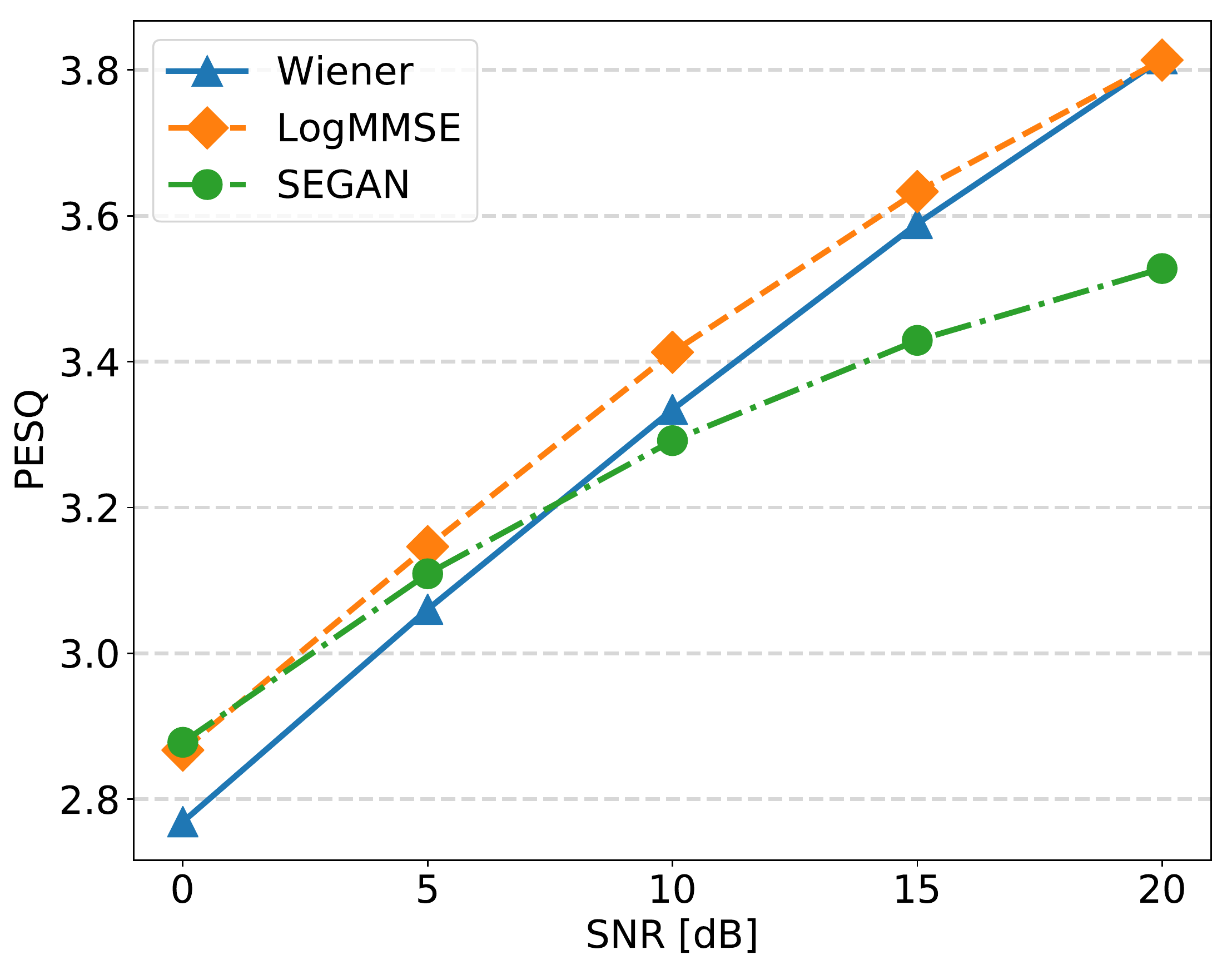}
\caption{\label{fig:pesq_avg}Average PESQ for different noise levels.}
\end{minipage}
\hfill
\begin{minipage}[c]{0.475\linewidth}
\includegraphics[width=\linewidth]{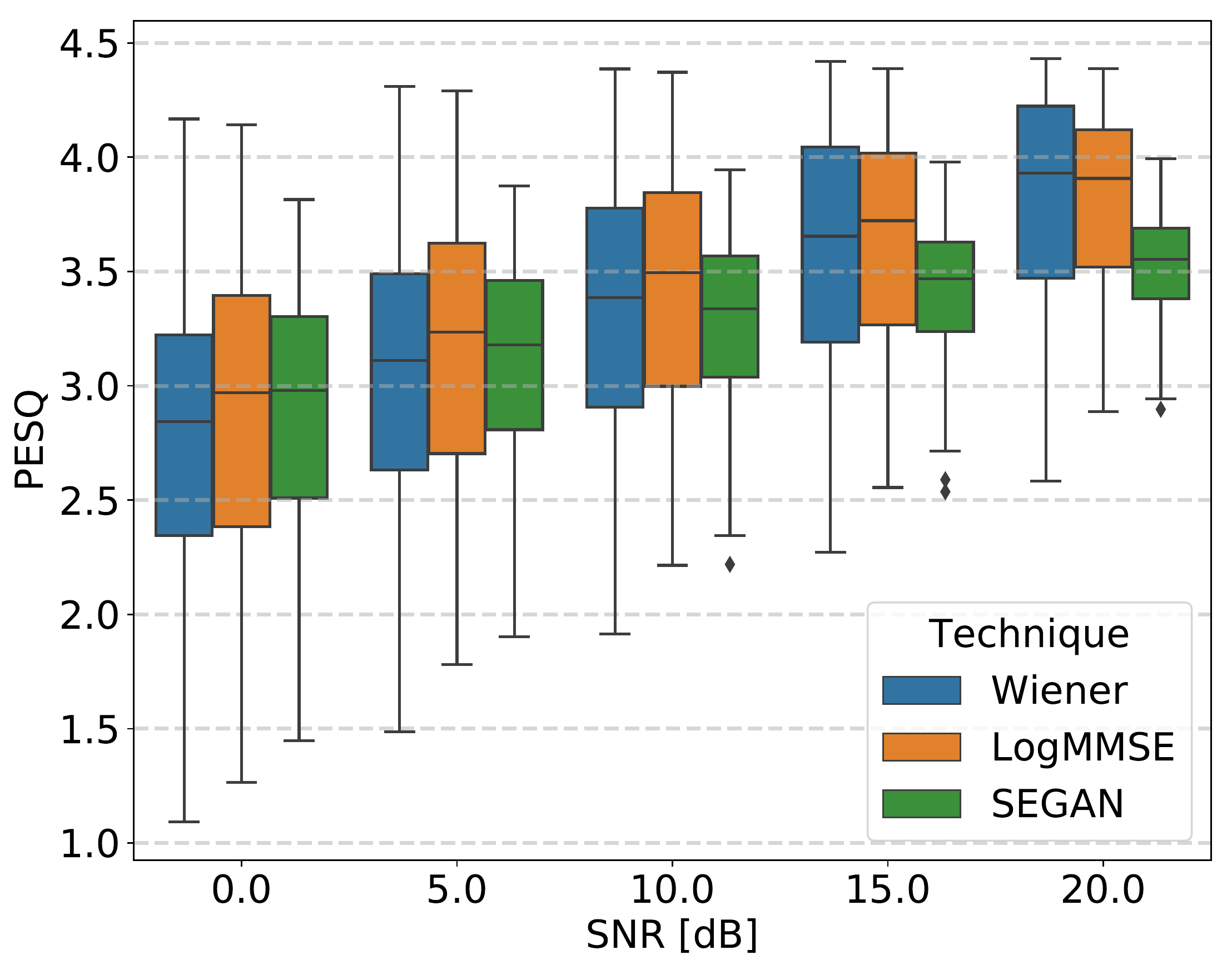}
\caption{\label{fig:pesq}PESQ for different noise levels.}
\end{minipage}%
\end{figure}

The average values of PESQ can be seen on Figures \ref{fig:pesq_avg} and \ref{fig:pesq}. It can be noted that the Wiener filter maintained a near linear behavior as the SNR was increased; although it was the technique with the worst performance for SNR 0 dB, it reached the best result of PESQ observed in this work for the case in 20 dB. LogMMSE has proven to be very effective from the start, going through all the scenarios as one of the best in terms of quality. SEGAN, on the other hand, showed a subtle superiority in the 0 dB scenario, but showed little improvement for higher SNR values, being on average much lower than the other two quality techniques analyzed during the enhancement process. Nevertheless, it is important to highlight the fact that the variance of SEGAN was lower in all scenarios.

\begin{figure}[htp]
\begin{minipage}[c]{0.475\linewidth}
\includegraphics[width=\linewidth]{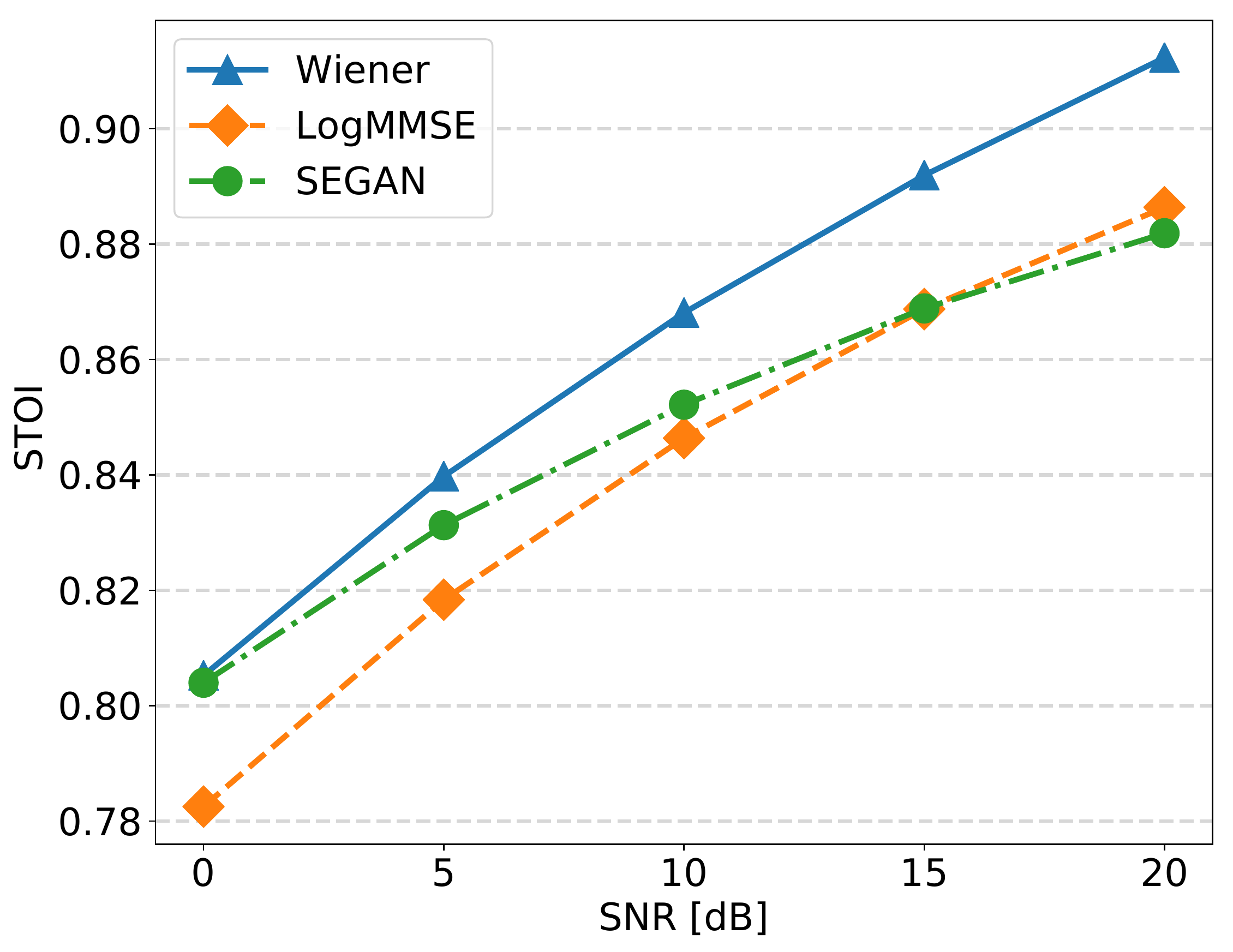}
\caption{\label{fig:stoi_avg}Average STOI for different noise levels.}
\end{minipage}
\hfill
\begin{minipage}[c]{0.475\linewidth}
\includegraphics[width=\linewidth]{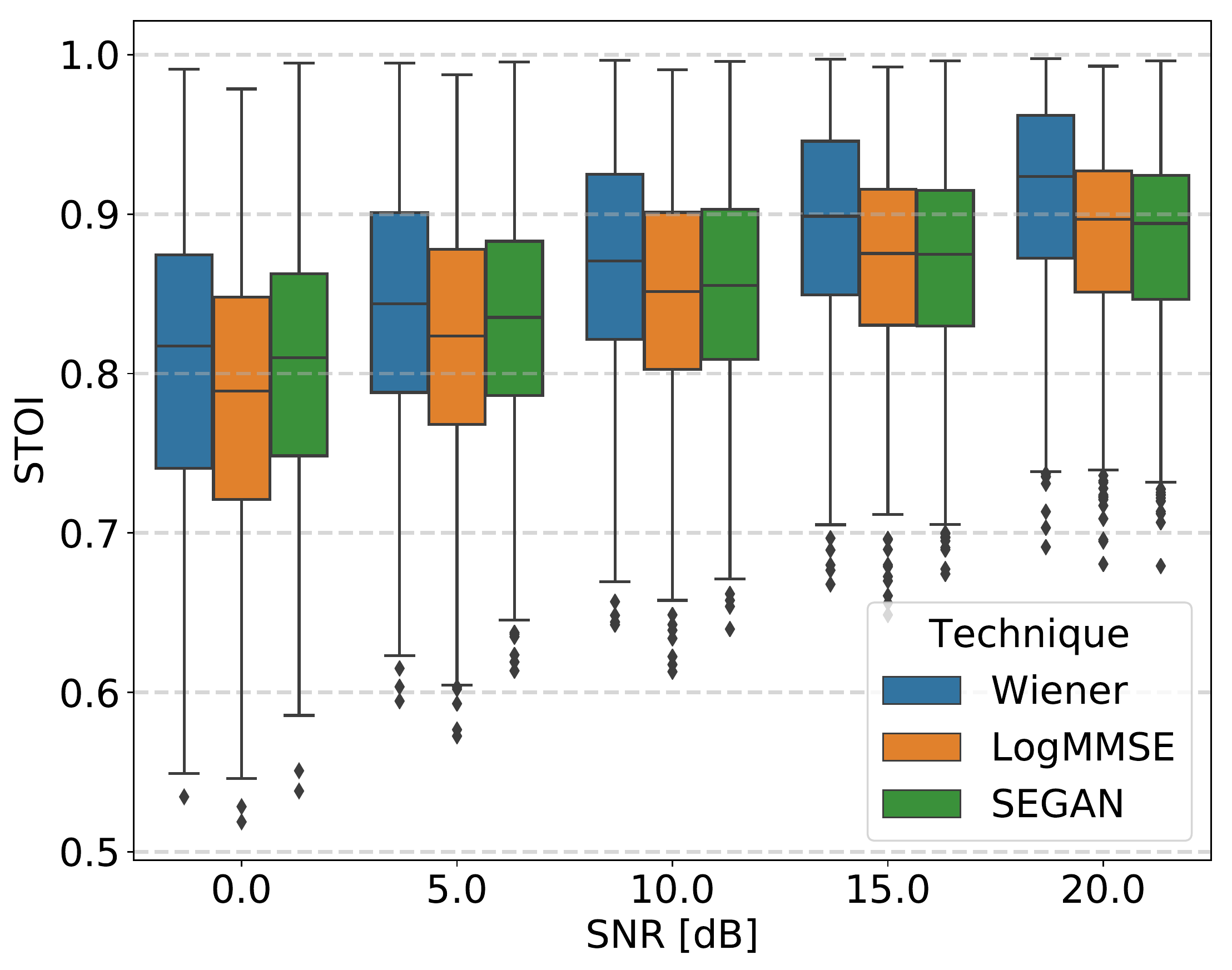}
\caption{\label{fig:stoi}STOI for different noise levels.}
\end{minipage}%
\end{figure}

Figures \ref{fig:stoi_avg} and \ref{fig:stoi} show STOI averages. The intelligibility is shown to be higher for the Wiener filter in all scenarios, which means that such a technique resulted in lower degeneration of speech comprehension. The LogMMSE showed a much lower performance than the other techniques for low SNR scenarios; improved slightly for 15 dB and 20 dB, but still got much worse than the Wiener filter. SEGAN showed a performance similar to that of the Wiener filter for the 0 dB scenario and remained superior to LogMMSE for this relation between the signal and distortions for almost all scenarios of different SNR values, except for 20 dB.

\begin{figure}[htp]
\begin{minipage}[c]{0.475\linewidth}
\includegraphics[width=\linewidth]{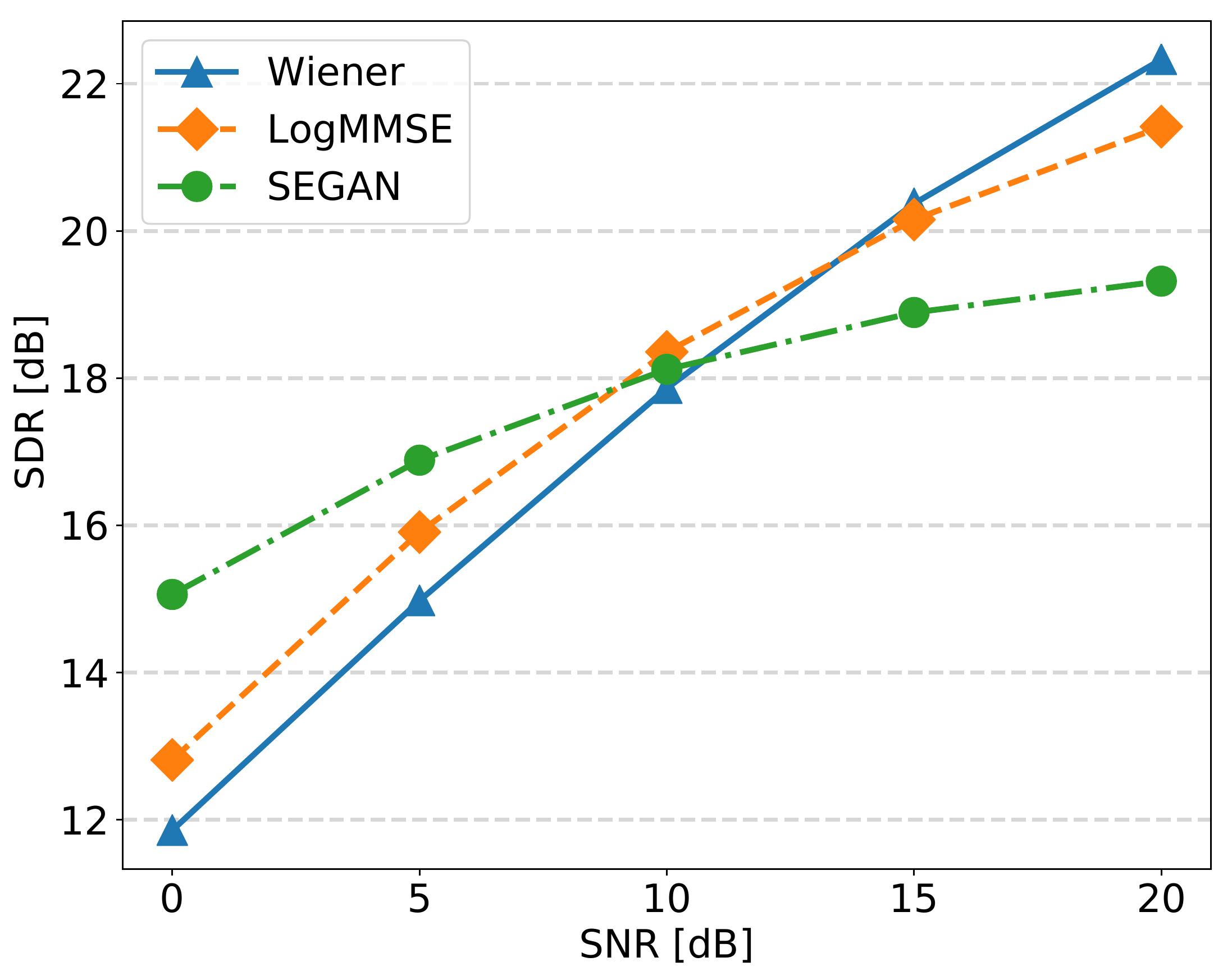}
\caption{\label{fig:sdr_avg}Average SDR for different noise levels.}
\end{minipage}
\hfill
\begin{minipage}[c]{0.475\linewidth}
\includegraphics[width=\linewidth]{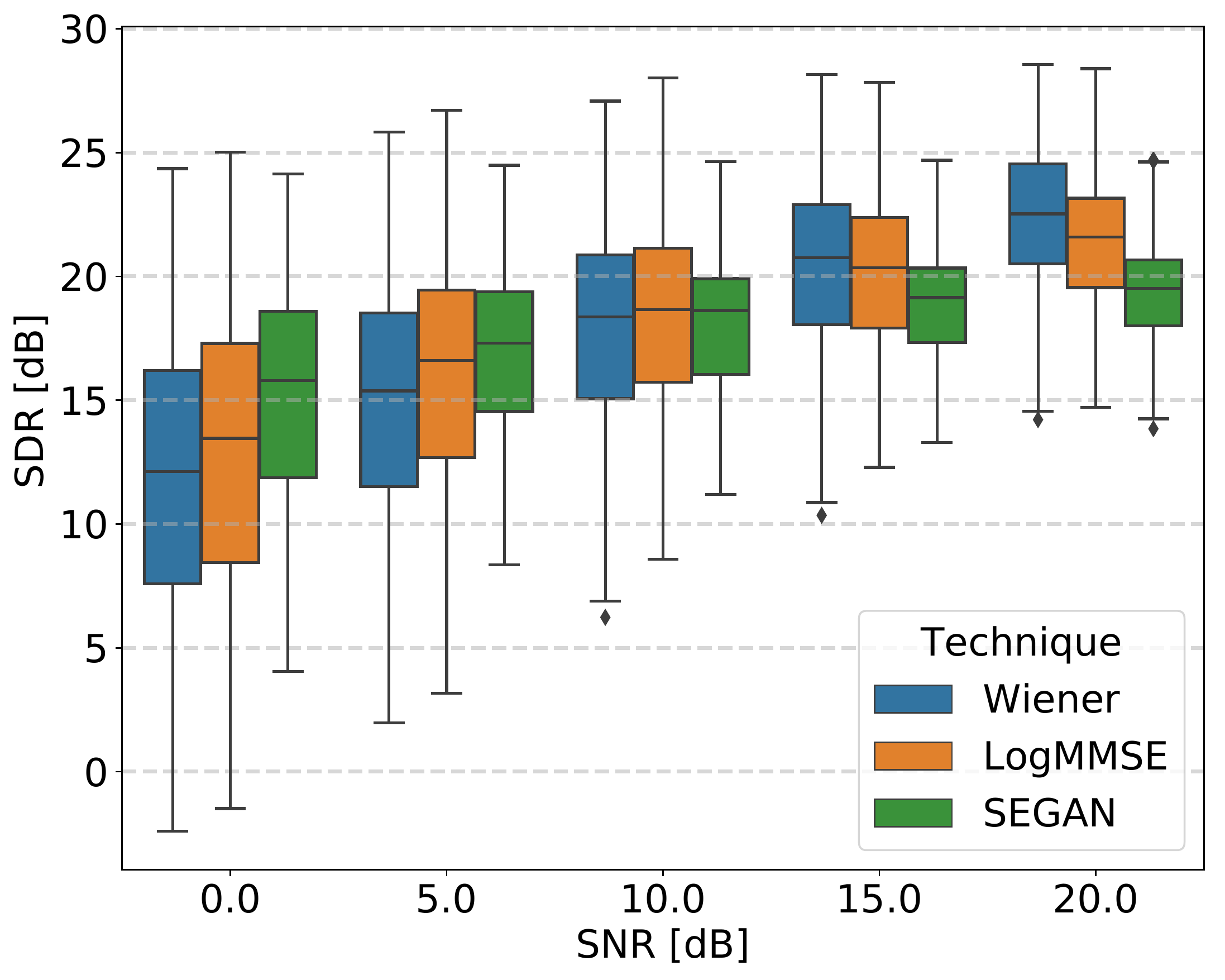}
\caption{\label{fig:sdr}SDR for different noise levels.}
\end{minipage}%
\end{figure}

The performance of each technique in terms of SDR can be seen in Figures \ref{fig:sdr_avg} and \ref{fig:sdr}. Notwithstanding the poor performance of Wiener filter for cases of lower SNR, it proved to be quite effective for higher SNR scenarios. The LogMMSE approach presented a similar performance to the Wiener filter. And, although it was the technique with the best performance for low SNR cases, SEGAN showed little improvement for cases with higher SNR; in the case with SNR 20 dB, its performance was well below that obtained by the other techniques; however, as with the PESQ metric, the variance of this technique was much lower than the other techniques.

\section{Discussion}

Although it was not the focus of this paper, there are some considerations to be made regarding the performance in terms of resources required for the speech enhancement process to be performed. While both classical methods adopted are based on an unsupervised approach, the GAN-based method is supervised, which requires a fundamental training step to be performed based on a pre-selected data set; and this step is computationally expensive as it took several hours to complete, even though it was performed on a GPU. Still, the application of the enhancement process itself through SEGAN's trained-model is not fast either, and it may take several seconds to complete the application over a single audio track of a few seconds. By contrast, classical methods performed the process almost instantaneously for each audio track and required no prior training.

Regarding speech quality and intelligibility, even considering the respectable, effective and efficient existing objective metrics, such as PESQ and STOI, respectively, if the enhanced signal is aimed for direct use by people, the use of metrics that still have the opinion of people, like \textit{Mean Opinion Score} (MOS) \cite{streijl2016mean}, may continue to be utilized, even if it has a lesser weight; after all, for various purposes, the human sense to evaluate and perceive distinct levels of quality may not yet have been well-enough designed in computational algorithms and metrics. The use of before-mentioned metrics would introduce factors of subjectivity into the process, which can be understood as something to be avoided; yet, if their influence is carefully managed in the appraisal and weighting, perhaps the results may be more satisfactory.

It is appropriate to indicate that Figure \ref{fig:stoi} also shows two critical details respecting the STOI results: an enormous variance and a colossal amount of outliers. The results show a considerable decrease in variance accompanied by a noticeable increase in already evident outliers. Such peculiarities described in this paragraph are worrisome regarding speech perception issues in noisy scenarios. This result may have been negatively affected by an innocently naive choice of complex audios, or by a training step that underwent from severe data frugality.

Given the observations indicated in the preceding paragraph, which emphasize certain undesired peculiarities about part of the results, especially regarding intelligibility, improvements may be perceived if meaningful arrangements are implemented to the assembled speech corpus. Perhaps using more numerous personalities of diverse ages, with different accents and more notable distinction in their vocal characteristics may enhance the results in future work.

About SDR, despite its relevance in this work, also because it is a usual objective metric, which tends to reduce human-related failures, it may be a less robust metric for some scenarios of speech enhancement or source separation, mainly for monoaural signals, which are of the type discussed in this paper. In order to address the problems associated with this metric, work \cite{le2019sdr} proposed an alternative metric called SI-SDR. Thus, in a future continuation of this work, this new metric proposal can be explored.

\section{Conclusions}

The results show that, although it is a classic technique confronted by more advanced ones, at least for the scenarios covered in this particular paper, the Wiener filter is still able to perform speech enhancement tasks for several scenarios, and remains a proper method for quality, intelligibility and distorting effects on speech signals. Despite its subtly inferior performance for some considered scenarios, SEGAN did well at 0 dB SNR scenarios, which are much more complicated, as well as exhibiting substantially lower variances. Although results obtained in \cite{SEGAN} indicate superiority over the Wiener filter, the divergence of results in relation to this work may be due to the wider variety of scenarios considered in this work. It is noteworthy, however, the need for a more detailed analysis of specific scenarios and also a more in-depth investigation into SEGAN.

\section*{Acknowledgment}

The authors are grateful for the support received from CAPES and from the Conselho Nacional de Desenvolvimento Científico e Tecnológico -- CNPq.

\bibliographystyle{unsrt}  
\bibliography{main.bib}  






\end{document}